\documentclass[fleqn,10pt]{wlscirep}
\usepackage[utf8]{inputenc}
\usepackage[T1]{fontenc}
\usepackage{amsmath}
\usepackage{amssymb}
\usepackage{braket}
\usepackage{graphicx}

\usepackage{verbatim}
\usepackage{float}

\usepackage{color}
\usepackage{soul}
\usepackage{siunitx}
\usepackage{gensymb}
\usepackage{bm}
\usepackage{xcolor}
\usepackage{multirow}

\title{Weak ferroelectric charge transfer in layer-asymmetric bilayers of 2D semiconductors}

\author[1,2]{F{\'a}bio Ferreira}
\author[1,2,3]{Vladimir V. Enaldiev}
\author[1,2,4]{Vladimir I. Fal'ko}
\author[1,2,*]{Samuel J. Magorrian}
\affil[1]{University of Manchester, Department of Physics \& Astronomy, Manchester, M13 9PL, United Kingdom}
\affil[2]{University of Manchester, National Graphene Institute, Manchester, M13 9PL, United Kingdom}
\affil[3]{Kotel'nikov Institute of Radio-engineering and Electronics of the Russian Academy of Sciences, Moscow, 125009, Russia}
\affil[4]{University of Manchester, Henry Royce Institute for Advanced Materials, Manchester, M13 9PL, United Kingdom}
\affil[*]{samuel.magorrian@manchester.ac.uk}

\begin{abstract}
In bilayers of two-dimensional (2D) semiconductors with stacking arrangements which lack inversion symmetry charge transfer between the layers due to layer-asymmetric interband hybridisation can generate a potential difference between the layers. We analyse bilayers of transition metal dichalcogenides (TMDs) - in particular, WSe$_2$ - for which we find a substantial stacking-dependent charge transfer, and InSe, for which the charge transfer is found to be negligibly small. The information obtained about TMDs is then used to map potentials generated by the interlayer charge transfer across the moiré superlattice in twistronic bilayers.
\end{abstract}
\begin{document}

\flushbottom
\maketitle
% * <john.hammersley@gmail.com> 2015-02-09T12:07:31.197Z:
%
%  Click the title above to edit the author information and abstract
%
\thispagestyle{empty}

\section*{Introduction}
Many two-dimensional (2D) materials\cite{Geim2013} lack inversion symmetry in their stoichiometric monolayers. These include hexagonal boron nitride (hBN), all TMDs, such as WSe$_2$, MoSe$_2$, etc., and the post-transition metal chalcogenides (InSe, GaSe). Depending on the orientation of the unit cells, bilayers of these materials can have inversion symmetry restored (like in 2H TMDs) or absent, like in $\gamma$-InSe. The orientation of the unit cells can therefore play an important role in determining electronic properties of twistronic structures of 2D semiconductors, where local stacking of the layers varies following the moiré superlattice structure. It is possible for layer-asymmetric structures to develop interlayer hybridisation between occupied (valence) and unoccupied (conduction) monolayer bands which leads to a net charge transfer between the layers\cite{Li2017, Tong2020, Woods2021, 2011.14237, 2010.05182, Sung2020, Enaldiev2021_hetero2DM}, resulting in a vertical electric field piercing the 2D crystal. In particular, for heterobilayers and nearly parallel (P-stacked) homobilayers of TMDs the charge transfer and resulting potential difference between the two layers will determine the features of band edge states of twisted bilayers.

Here, we focus on the theoretical modelling aspect of ferroelectric charge transfer in 2D materials. The potential variation across a layer-asymmetric 2D material system in which charge transfer has occurred is a natural subject for first principles density functional theory (DFT) calculations.
In modelling a 2D system using a plane-wave DFT code, to satisfy the requirement for periodicity in the out-of-plane direction, it is typical to construct a supercell such that repeated images of the few-layer crystal are separated by a large vacuum, minimising the interaction between them.
Charge transfer will give the electrostatic vacuum potentials on either side of the 2D system different values, violating the requirement for out-of-plane periodicity in the DFT calculations.
In this study, we compare results which do not correct for the effect of a polar bilayer with calculations using two means by which periodicity is often satisfied - first the construction of a supercell containing two images of the system, one with the layers interchanged, second the approximate method of applying a compensating step potential in the vacuum region - a `dipole correction'\cite{Dipole_1, Dipole_2}.

This article is structured as follows. First we give a full account of the DFT methods used in this work, with WSe$_2$ employed as a test system.
Next, we explore charge transfer effects across a range of semiconducting TMDs.
Then, we analyse InSe bilayers and show that they exhibit much weaker interlayer charge transfer as compared to WSe$_2$, WS$_2$, MoSe$_2$ and MoS$_2$.
Finally, we implement information collected about stacking-dependent interlayer charge transfer in TMD bilayers to discuss its manifestation in the domain structure of twistronic TMDs.

\section*{DFT calculations of interlayer charge transfer in semiconductor bilayers}
\label{sec:DFT}

\subsection*{Achieving out-of-plane periodicity}
Density Functional Theory (DFT) calculations of ultrathin films of layered materials using plane-wave based methods require the resolution of a crystal which is only periodic into two dimensions into a three-dimensionally periodic system. This is usually achieved by the construction of a supercell in which images of the two-dimensional (2D) layers are repeated periodically along the third dimension, with a large vacuum between them to ensure the layers are isolated from each other. Here, we compare results for polar WSe$_2$ bilayers using three commonly-used methods of addressing the requirement for periodicity in first-principles calculations of 2D materials: (i) a supercell with a single bilayer\cite{Sung2020}, (ii) a single bilayer supercell, but with a dipole correction applied\cite{PhysRevB.99.075160}, and (iii) a supercell with two mirror-reflected images of the bilayer  \cite{Tong2020}.

We consider these methods in XM$'$-stacked WSe$_2$ (the prime symbol indicating that of the vertical metal-chalcogen pair in the bilayer, the metal atom is in the top layer\cite{Enaldiev_arXiv}), the structure of which we show in Fig. 1. %\ref{fig:structure}.
In the upper panel of Fig. 2%\ref{fig:pots}
, we show its plane-averaged local electrostatic potential (ionic and Hartree contributions) relative to the vacuum level on the Se side of the vertical W-Se pair, which is set to 0~eV. While the detailed variation of the local potential in the vicinity of the atomic planes depends sensitively on how the local and non-local parts of the pseudopotentials are set up, an important meaningful quantity can be extracted from the difference between the vacuum levels on both sides. As modelled in a previous work\cite{Enaldiev2021_hetero2DM}, layer-asymmetric hybridisation between occupied and unoccupied states gives rise to charge transfer between the layers, giving the bilayer a finite dipole moment which results in the vacuum potentials on either side of the slab having different values. As shown schematically as an inset, the calculation was carried out using a supercell containing two mirror images of the bilayer (the second image thus being MX$'$-stacked), separated by large vacuums. Then, the requirement for out-of-plane periodicity in a plane wave DFT code is met, with the vacuum potentials matching at the supercell boundary.

In the left-hand lower panel of Fig. 2%\ref{fig:pots}
, we compare the results for the first bilayer of the double-bilayer supercell with two methods from a supercell containing a single bilayer.
To more easily see the effects of interlayer charge transfer, we subtract the potentials of isolated monolayers from the bilayer potentials.
In a supercell with only one bilayer, the periodicity in the out-of-plane direction is broken, and there is a mismatch between the vacuum potentials at the supercell boundary.
The numerical result of this in the DFT calculation is an artificial displacement field experienced across the slab, as can be seen in the red line in Fig. 2. %\ref{fig:pots}.
The difference between the single- and double-bilayer supercells is greatest in the vacuum regions, as the effect of the artificial field is partially suppressed by the dielectric response within the WSe$_2$ bilayer itself.

We also show a further set of results for a single supercell, but with the well-known dipole correction applied\cite{Dipole_1, Dipole_2}. This adds a background sawtooth potential at each electronic self-consistency step of the calculation, with its size calculated from the electric dipole moment of the bilayer.
As can be seen from the agreement between the black and blue lines in Fig. 2, %\ref{fig:pots}
this is a very good approximation to the truly periodic double-bilayer, while computationally much cheaper, being a supercell of half the size. In Table 1, %\ref{tab_fit_splitting}
we show a few key energies from the band structures resulting from the various supercell methods. While the single- and double-bilayer supercell methods give slightly different results, the application of a dipole correction to the single-bilayer supercell returns the values to good agreement with those for the double-bilayer supercell.

\subsection*{Choosing DFT functional}
For comparision with the PBE GGA results presented above, calculations were also carried out using the local density approximation (LDA), via the exchange correlation functional of Ceperley and Alder\cite{PhysRevLett.45.566} as parametrized by Perdew and Zunger\cite{PhysRevB.23.5048}, and in Fig. 2 %\ref{fig:pots}
we show the differences between bilayer and monolayer plane-averaged local electrostatic potentials using LDA alongside the PBE results. In Table 1 %\ref{tab_fit_splitting}
the LDA-calculated potential differences and some band energies are compared with the PBE results. The bilayer-monolayer potential differences show a drop across the bilayer similar to that seen using PBE, but with a notable peak in between the layers. The potential difference across the bilayer, and the splitting between the upper K-point valence bands show small differences of only a few meV, but the difference between the $\Gamma$- and K-point valence band energies is notably reduced on going from the PBE approximation to LDA. Since the pseudopotential configurations as shown in Table 2 are nearly identical, we ascribe the differences between the calculations to the approximation to the exchange-correlation potential used.

\section*{Configuration dependence of weak ferroelectric effect in P-stacked bilayers}
\label{sec:TMD}
We now study various TMD bilayers with XM$'$/MX$'$ configuration as well as a number of various stackings which will enable us to describe the variation of charge transfer across the moiré supercell of a twisted bilayer. The strength of hybridisation between the layers is sensitive to the sublattice composition of the band states, and it is stronger for states residing on the chalcogen sublayers.
In Table 3 %\ref{tab:wfns_four_tmds}
we compare (with calculations using the QE code and a double-bilayer supercell) the wavefunction projections onto the six atomic layers of MX$'$ stacked TMD bilayers. As noted previously for MoSe$_2$ bilayers\cite{Sung2020}, but repeated in a manner common to all four TMDs considered, the K-point wavefunctions are nearly entirely layer-polarised, due to a combination of very weak interlayer intraband hybridisation and the effective electric field between the layers arising from the charge transfer effect discussed above. In contrast, the $\Gamma$-point valence band wavefunction has an almost zero out-of-plane dipole moment, due to the strong interlayer hybridisation of the $\Gamma$-point states. The Q-point, which in some cases forms the conduction band minimum, is an intermediate case.

The variation of the potential drop across a P-MX$_2$ bilayer with in-plane offset, $\bm{r}_0$ ($\bm{r}_0=0$ for XX-stacking corresponding to overlapping of chalcogens in two layers), and interlayer distance, $d$, can be described using the following expression\cite{Enaldiev2021_hetero2DM},
\begin{equation}\label{Eq:potential_jump}
\Delta^{P}(\bm{r}_0,d) = \Delta_a e^{-q(d-d_0)}\sum_{j=1,2,3}\sin(\bm{G}_j\cdot\bm{r_0}),
\end{equation}
where values of parameters $\Delta_a$, $q$ and $d_0$, $\bm{G}_{1,2,3}$ are the shortest reciprocal vectors of a monolayer related by $120^{\circ}$-rotation around $z$. This formula is applicable to all TMDs with a honeycomb lattice structure, with parameters for MoS$_2$, MoSe$_2$, WS$_2$ and WSe$_2$ calculated using the QE code, shown in Table 4. %\ref{tab_potential drops_fit}.
Since in twisted bilayers the interlayer distance $d$ and in-plane offset $\mathbf{r}_0$ vary continuously, we will use the information presented here concerning the dependence of charge transfer on stacking configuration to map the charge transfer and on-layer potential in a moiré superlattice.

\section*{Indium Selenide}
\label{sec:InSe}

As a comparison to the transition metal dichalcogenides, we consider indium selenide, a member of the family of post-transition metal chalcogenides. The two most commonly found polytypes of bulk InSe in experiments are the $\gamma$\cite{Rigoult1980} and $\varepsilon$\cite{Grimaldi2020} polytypes - on exfoliation to a bilayer, these will both have the same layer-asymmetric MX$'$/XM$'$ character to their stacking order. In the two panels of Fig. 3 %\ref{fig:pots_InSe}
we show first the local electrostatic potentials with isolated monolayer contributions subtracted comparing the supercell and dipole correction methods, and in the second the same differences, but calculated using LDA. For InSe, the peak in the difference between monolayer and bilayer potentials in the interlayer region shown in the LDA results for WSe$_2$ is present for both PBE and LDA approximations - but is much greater in magnitude for the LDA case. The charge transfer for bilayer InSe is negligible, giving a difference of only $\sim$2~meV between the vacuum potentials across the bilayer, with the consequence that differences between supercell and correction methods are negligible.
\section*{Mapping charge transfer across the moiré superlattice of twistronic bilayers}
\label{sec:Conclusions}

While the asymmetry of band-edge states in inversion asymmetric bilayers manifests itself in the linear Stark shift of band energies and of the energies of optical transitions \cite{Sung2020} in vertically biased MX$'$ and XM$'$ bilayers, the ferroelectric potentials can be detected by contrast in the potential maps of systems with laterally varying stacking. Such a variation naturally appears in twisted bilayers with a parallel orientation of monolayer unit cells, where the twist angle determines the period of recurrent stacking configurations, known as moiré superlattice. To describe such variation, we employ the description of interlayer potential and the related size of the double layer of charge, $\pm \delta \rho$, on the top/bottom layers.

To demonstrate the latter, in Fig. 4 %\ref{fig:charge_diff}
we show the $z$-coordinate dependence of the difference between the plane-averaged charge density of XM$'$ stacked bilayer WSe$_2$ and that of two isolated WSe$_2$ monolayers. The greatest differences and charge transfer are to be found in the interlayer region, with a peak and a trough close to the inner Se atomic layers. This peak(trough) could be related to the (de)population of hybridised $s$ and $p_z$ orbitals on the Se atoms.
We can calculate a charge transfer density directly from DFT as
\begin{equation}
  \delta\rho = \frac{1}{2}\left(\int_0^{15}\rho(z)dz - \int_{15}^{30}\rho(z)dz\right),
\end{equation}
where $\rho(z)$ is the plane-averaged charge density at the $z$-coordinates as shown in Figs.~2~and~4, and $z=15$~\AA~ is the mean plane between the two WSe$_2$ layers. This gives $\delta\rho = 1.9\times 10^{12}$~cm$^{-2}$.

We can also roughly estimate the magnitude of charge density transferred between the layers based on the potential drop, $\Delta^P$ (Eq. \eqref{Eq:potential_jump}), as
    $\dfrac{\varepsilon_0\Delta^P}{e^2d_{XX}}$
where $d_{XX}$ is the distance between the inner chalcogen atomic planes (3.14~\AA~for WSe$_2$). This gives $\sim 10^{12}$~cm$^{-2}$ for MX$'$/XM$'$ stacked bilayer WSe$_2$, with the net transfer of electrons being to the layer in which the selenium atoms are vertically opposite the tungsten atoms in the other layer. The difference in the precise numerical values between that directly calculated from DFT, and that from the rough estimate, arises as the greatest part of the charge transfer occurs in the interlayer region, over a smaller distance than the $d_{XX}$ used to convert from the potential drop.

In a rigid twisted bilayer (which corresponds to $\theta_P > 2.5^{\circ}$\cite{Enaldiev_arXiv}) the relative in-plane shift of the layers is $\bm{r}_0 = \theta\hat{z}\times \bm{r}$. In a reconstructed twistronic bilayer of a marginally twisted ($\theta_P\ll 1^{\circ}$) parallel(P)-stacked TMD\cite{Enaldiev_arXiv, Weston2020}, a pattern of triangular domains forms, with alternating  MX$'$/XM$'$ stacking, with $\bm{r}_0 = \theta\hat{z}\times \bm{r} + \bm{u}(\bm{r})$, where $\bm{u}(\bm{r})$ is the relative displacement field of the two layers, formed on reconstruction. Within this domain pattern, there will be an excess of bonded electron charges in the top layer in the centre of MX$'$ domains, and a corresponding deficiency in XM$'$ domains, with a general $\bm{r}_0,d$ dependence in the resulting potential distribution given by Eq. \eqref{Eq:potential_jump}. An estimate of the magnitude of the charge density transferred between the layers can be roughly approximated as set out above.

The distribution of potential above a twistronic bilayer resulting from its out-of-plane polarisation can be mapped using scanning Kelvin probe microscopy \cite{Nonnenmacher1991}(SKPM) or a single electron transistor \cite{Yoo1997, Martin2007}.
The SKPM signal would vary between MX$'$ and XM$'$ regions, with the magnitude of variation dependent on the distance from the scanning tip to the surface, and between the bilayer and metallic back plate, which provides the reference for the locally measured potential.
For a structure with a thick dielectric substrate separating the bilayer from the back plate by more than the superlattice period, as shown schematically in Fig. 5a, the potential measured by the tip close to the bilayer surface would display a variation with amplitude $\Delta^P({\rm MX'})$, as shown on the map.
For a structure where the back plate is at a distance from the bilayer much less than the moir{\'e} superlattice period (Fig. 5b), the potential variation between XM$'$ and MX$'$ domains would be $2\Delta^P({\rm MX'})$. The corresponding values of $\Delta^P$ are listed in Table 4.
Subject to the requirement that the bilayer remains undoped (so that lateral potential screening inside the bilayer would not kill the effect), the described behaviour should be expected in all twisted TMD bilayers discussed in this work, as well as in heterobilayers\cite{Enaldiev2021_hetero2DM}.

In conclusion, we have described interlayer charge-transfer effects in 2D semiconductor bilayers. Two means of maintaining the requirement of out-of-plane periodicity in DFT calculations have been discussed. We have shown a substantial effect in the TMDs, demonstrating a general formula for finding the size of the effect for general lattice configurations beyond commonly-found high-symmetry stacking orders, showing how charge transfer will be important in understanding the behavior of twistronic 2D material structures, where the local atomic registry and interlayer distance will vary continuously. In contrast to the TMDs, the size of the charge transfer effect in the hexagonal post-transition metal chalcogenide InSe is found to be negligibly small.

\section*{Methods}

The DFT calculations for WSe$_2$ and InSe in this work were carried out using the plane-wave based VASP code \cite{VASP} using the projector augmented
wave (PAW) pseudopotentials as distributed with VASP 5.4.4\cite{PhysRevB.50.17953, PhysRevB.59.1758}. We approximated the exchange correlation functional using the generalised gradient approximation (GGA) of Perdew, Burke and Ernzerhof (PBE method)\cite{PBE}, while for the local density approximation (LDA) comparisons in Figs. 1 and 2 we use the exchange correlation functional of Ceperley and Alder\cite{PhysRevLett.45.566} as parametrized by Perdew and Zunger\cite{PhysRevB.23.5048}. In Table 2 we give cutoff radii and valence electron configurations for the VASP pseudopotentials used.
The cutoff energy for the plane-waves is set to 600 eV with the in-plane Brillouin zone sampled by a
$12 \times 12$ grid.  Monolayer crystal structure parameters and interlayer distances are taken from experimental references for bulk crystals\cite{schutte1987crystal, bronsema1986structure, Rigoult1980}.

We also compare results for four of the TMDs using calculations carried out using the Quantum Espresso (QE) package \cite{QE1,QE2}.
A plane-wave cutoff energy of 1090 eV was used for all QE calculations, where the integration over the Brillouin zone was performed
using scheme proposed by Monkhorst-Pack with a grid of $12 \times 12\times 1$.
All calculations used full relativistic norm-conserving pseudopotentials with spin-orbit interaction included. The exchange correlation functional was approximated using the PBE method.

\section*{Data availability}
Modelling inputs and resulting data generated in this study are available from the authors on reasonable request.

\section*{Acknowledgements}

We thank D. A. Ruiz-Tijerina, N. Walet, R. Gorbachev, A. K. Geim, A. Weston, Q. Tong, M. Chen, F. Xiao, H. Yu, W. Yao, and V. Z{\'o}lyomi for discussions.
This work has been supported by EPSRC grants EP/S019367/1, EP/S030719/1, EP/N010345/1, EP/V007033/1; ERC Synergy Grant Hetero2D; Lloyd’s Register Foundation Nanotechnology grant; European Graphene Flagship Project, and EU Quantum Technology Flagship project 2D-SIPC. Computational resources were provided by the Computational Shared Facility of the University of Manchester, and the ARCHER2 UK National Supercomputing Service (https://www.archer2.ac.uk) through EPSRC Access to HPC project e672.

\section*{Author contributions statement}
S.J.M. and V.F. conceived the project, F.F. and S.J.M. carried out DFT calculations. All authors contributed to the analysis of the DFT data for developing models applicable to twistronic structures, V.V.E. and F.F. modelled lattice reconstruction. All authors contributed to writing the manuscript.

\section*{Additional information}

\textbf{Competing interests} The authors declare no competing interests.

\begin{figure}[ht]
    \centering
    \includegraphics[width = 0.8\linewidth]{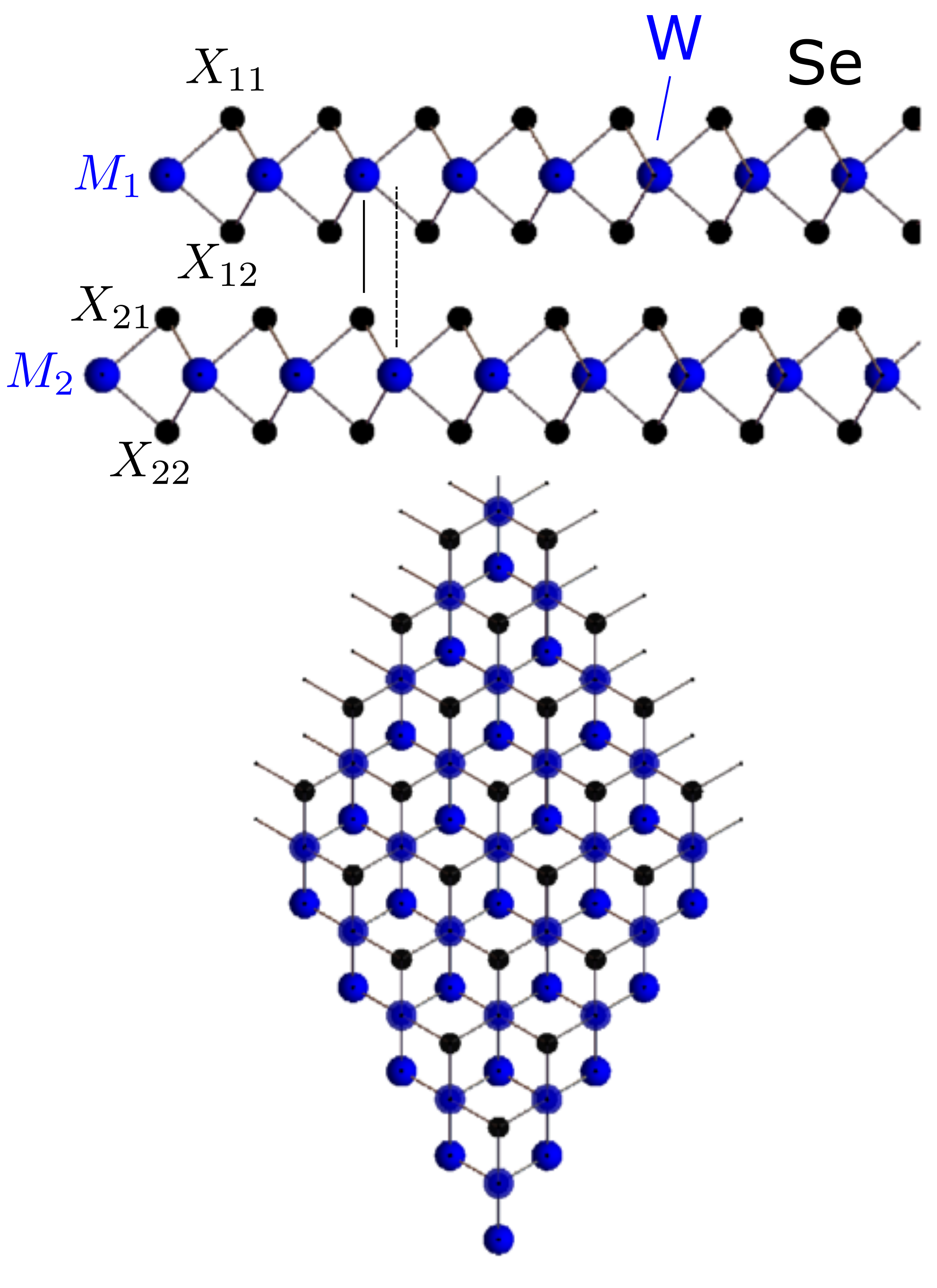}
    \caption{Upper and lower panels: top and side view, respectively, of XM$'$-stacked WSe$_2$ (the prime indicating that of the vertical metal(M)-chalcogen(X) pair, the metal atom is in the top layer). The solid and dashed vertical lines in the upper panel emphasise the symmetry-breaking of this stacking, with the metal-chalcogen pair vertically opposite in one direction, but not in the other. The italicised labels in the upper panel ($M_1, X_{12}$ etc.) indicate the atomic positions referred to in Table 3.}
    \label{fig:structure}
\end{figure}

\begin{figure}[ht]
    \centering
    \includegraphics[width = 0.7\linewidth]{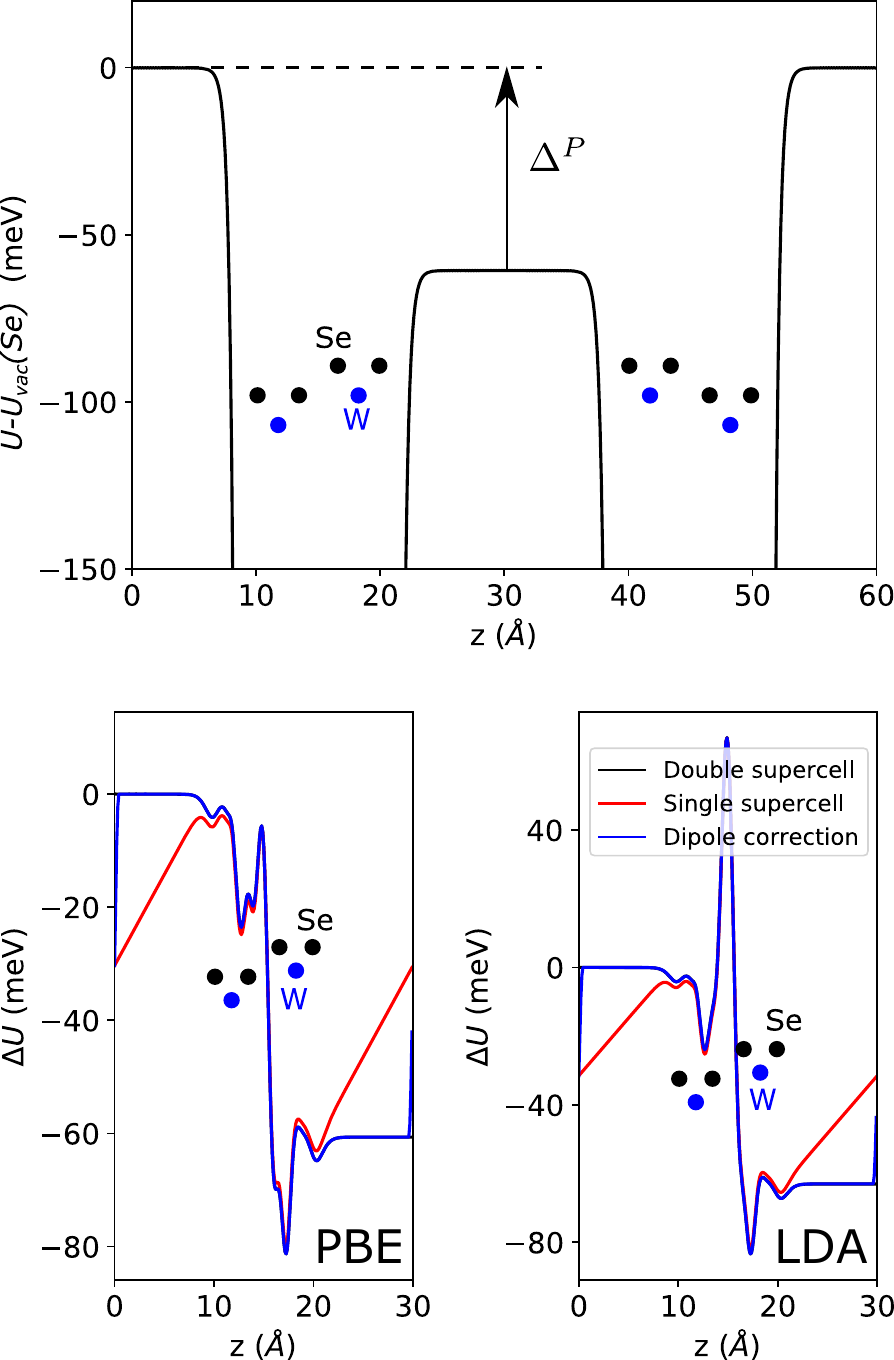}
    \caption{Upper panel: Out-of-plane dependence of in-plane averaged electron potential energy of XM$'$ bilayer WSe$_2$ (including ionic and Hartree contributions), relative to the vacuum potential on the Se side of the vertical W-Se pair, $U_{vac}(Se)$, which is set to 0~eV. The calculation is for a double-bilayer supercell, with the structure shown as a schematic inset. The charge transfer between the layers gives each bilayer a finite dipole moment, with a consequent difference ($\Delta^P$) between the vacuum levels on either side of a bilayer. Left-hand lower panel: potentials with isolated monolayer contributions subtracted, to better show the potential drop and other features. Three calculation methods are shown: black line - first 30~\AA~ of the upper panel, red line - using only a single bilayer supercell, showing how the mismatched vacuum levels give a  finite displacement field as necessitated by the periodic boundary conditions, blue line - a single bilayer supercell, but with a compensating dipole correction applied. The right-hand lower panel shows the same quantities, calculated using the LDA for comparison. 0~meV is set to the double-supercell vacuum level on the Se side of the vertical W-Se pair.}
    \label{fig:pots}
\end{figure}

\begin{figure}[ht]
    \centering
    \includegraphics[width = \linewidth]{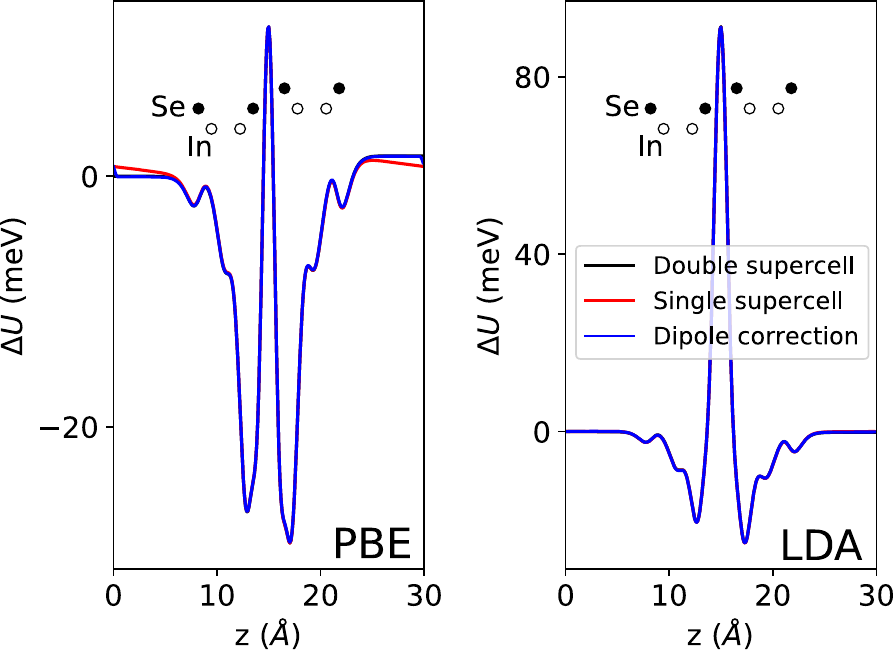}
    \caption{Left panel: difference between bilayer and isolated monolayer plane-averaged local potentials for 3R-stacked bilayer InSe, comparison between supercell methods. Right panel: same as left panel, but calculated using LDA. 0~meV is set to the double-supercell vacuum level on the Se side of the vertical In-Se pair.}
    \label{fig:pots_InSe}
\end{figure}

\begin{figure}[ht]
    \centering
    \includegraphics[width=\linewidth]{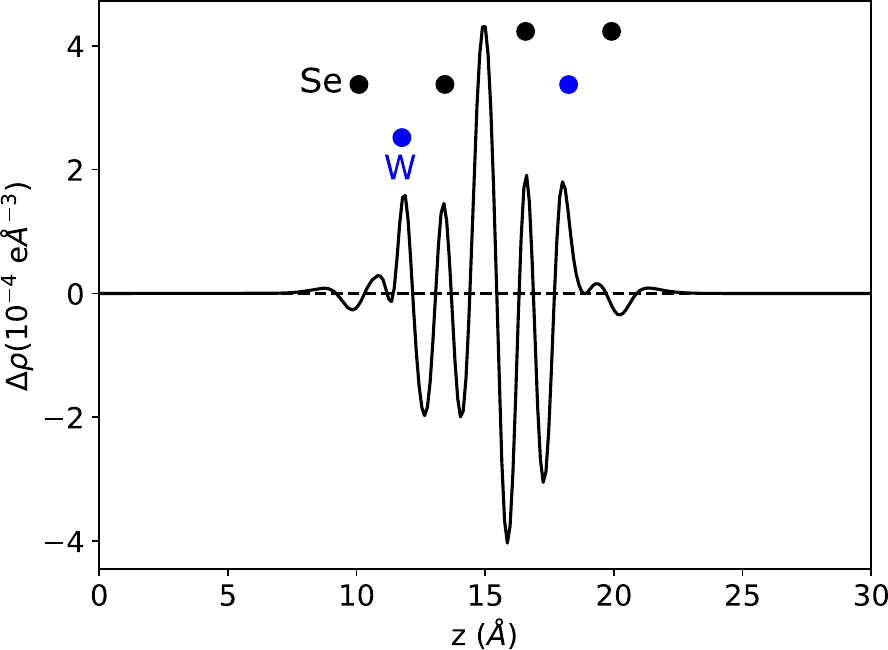}
    \caption{Difference between plane-averaged charge density of an XM$'$ bilayer and two isolated monolayers, showing greatest charge transfer in interlayer region between planes of Se atoms. Dashed line is a guide to the eye, showing $\Delta\rho = 0$.}
    \label{fig:charge_diff}
\end{figure}

\begin{figure}[ht]
    \centering
    \includegraphics[width = 0.9\linewidth]{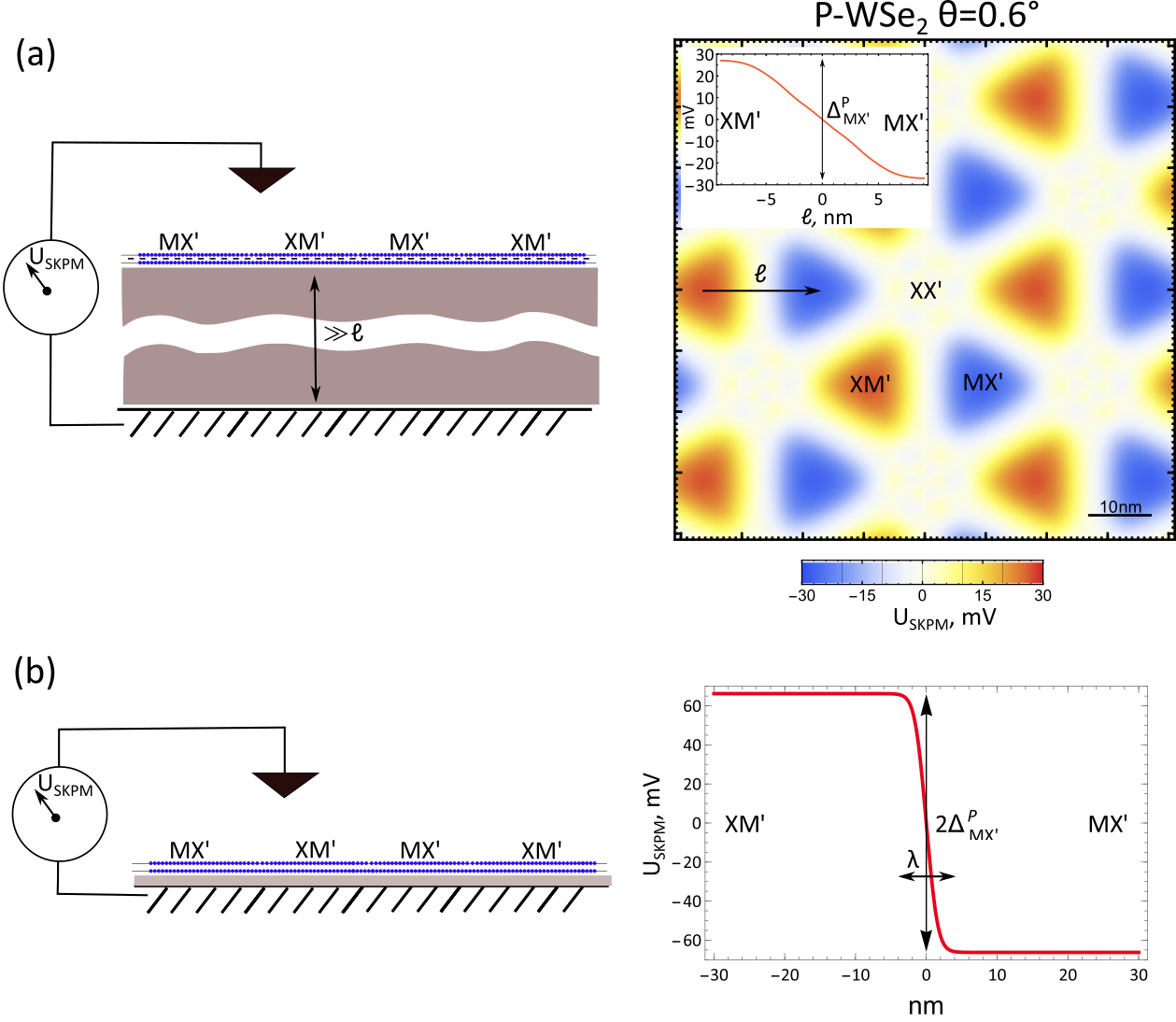}
    \caption{(a) Left panel: schematic of Kelvin probe setup where TMD bilayer is placed on dielectric substrate with thickness much larger than moir\'e period $\propto \ell$. Potential differences are then measured w.r.t. the middle plane of the TMD bilayer, with the resulting patterns of potential exemplified in right panel: map of top layer potential of a nearly parallel-stacked WSe$_2$ bilayer with twist angle $\theta=0.6^{\circ}$, from sum of piezoelectric potential\cite{Enaldiev_arXiv} contribution and the charge transfer described in this work. Inset: shape of potential drop on going from an XM$'$ domain to MX$'$ (path marked with an arrow in the map), with an in-plane electric field at the domain wall. (b) Left panel: for a twisted bilayer sample placed separated from a  metallic plate by only a very thin dielectric of thickness much less than the moir{\'e} superlattice period, the potential in the bottom layer will be the same for all domains. This will double the potential difference measured across a domain wall in the top layer, shown in the right panel. The drop of potential occurs on the length $\lambda\approx 8$\,nm corresponding to the width of the domain wall\cite{Enaldiev_arXiv}.}
    \label{fig:potential_map}
\end{figure}

{%\setlength{\tabcolsep}{0.1\columnwidth}%10.2pt}
\setlength\extrarowheight{2pt}
\begin{table}
	\caption{
		{Quantities (meV) calculated using different supercell methods for XM$'$ bilayer WSe$_2$ - 2$\times$(1$\times$)BL: double(single) bilayer supercell), Dip.: Dipole-corrected single bilayer supercell.  $\Delta^P$ is the difference between vacuum energies, while $E_{VB,K} - E_{VB-1,K}$,
		$E_{VB,K} - E_{VB,\Gamma}$, and
		$E_{CB,K} - E_{VB,K}$ are the splitting between the top the valence bands at the K-point, the difference between the local $\Gamma$- and K-point valence band maxima, and the vertical K-K gap, respectively. \label{tab_fit_splitting}
		}
	}
	\begin{tabular}{cccc}
		\hline\hline
		&  2$\times$BL & 1$\times$BL & Dip. \\
		\hline
		PBE &\multicolumn{3}{c}{Energies in meV}\\
		%\hline
		$\Delta^P$ & 61  &n.a. & 61 \\
		$E_{VB,K} - E_{VB-1,K}$ & 70 & 67 & 70 \\
		$E_{VB,K} - E_{VB,\Gamma}$ & 80 & 79 & 80\\
		$E_{CB,K} - E_{VB,K}$ & 1289 & 1291 & 1289\\
		\hline
		LDA & & & \\
		$\Delta^P$ & 63  &n.a. & 63 \\
		$E_{VB,K} - E_{VB-1,K}$ & 73 & 70 & 73 \\
		$E_{VB,K} - E_{VB,\Gamma}$ & 60  & 59 & 60\\
		$E_{CB,K} - E_{VB,K}$ & 1277 & 1280 & 1277\\
		\hline
		\hline
	\end{tabular}
\end{table}
}

\begin{table}[ht]
	\caption{
		{Details of projector augmented wave (PAW) pseudopotentials used in VASP calculations of WSe$_2$ and InSe. The date identifies the generation of the specific pseudopotential. Config. gives the valence electron configuration of the calculation. RMAX is the core radius for the PAW projector operator, RDEPT is the core radius for the augmentation charge, and RCUT is the core cutoff radius.  \label{tab:pseudops}
		}
	}
	\begin{tabular}{c|cc|cc|cc}
		\hline\hline
		&  W (PBE) & W (LDA) & In (PBE) & In (LDA) & Se (PBE) & Se (LDA)\\
		\hline
		Date &08Apr02&19Jan01&08Apr02&03Oct01&06Sep00&03Mar98 \\
		Config. &5d$^4$6s$^2$&5d$^4$6s$^2$&5s$^2$5p$^1$&5s$^2$5p$^1$&4s$^2$4p$^4$&4s$^2$4p$^4$ \\
		RMAX &1.482~\AA&1.482~\AA&1.676~\AA&1.676~\AA&1.136~\AA&1.136~\AA \\
		RDEPT &1.180~\AA&1.175~\AA&1.435~\AA&1.435~\AA&1.021~\AA&1.012~\AA \\
		RCUT &1.320~\AA ($l=0,2$)&1.320~\AA ($l=0,2$)&1.640~\AA&1.640~\AA&1.110~\AA&1.110~\AA \\
		&1.455~\AA ($l=1$)&1.455~\AA ($l=1$)&&&&\\
		\hline
		\hline
	\end{tabular}
\end{table}

\begin{table}[ht]
\caption{Wavefunction projections onto atomic layers for XM$'$ stacked TMD bilayers, for atomic positions shown in Fig. 1, together with the dipole moment $d_z = \bra{\psi}z\ket{\psi}$, for valence (VB) and conduction (CB) states at important points in the Brillouin zone (Q=K/2).\label{tab:wfns_four_tmds}}
\resizebox{\columnwidth}{!}{%
\begin{tabular}{l|cccc|cccc|cccc|cccc}
\hline\hline & \multicolumn{4}{c|}{MoS$_2$} & \multicolumn{4}{c|}{MoSe$_2$} & \multicolumn{4}{c|}{WS$_2$} & \multicolumn{4}{c}{WSe$_2$} \\
\hline$\left| \psi \right|^2$ (\%)     & K CB   & Q CB   & K VB  & $\mathrm{\Gamma}$ VB & K CB   & Q CB   & K VB  & $\mathrm{\Gamma}$ VB & K CB   & Q CB   & K VB  & $\mathrm{\Gamma}$ VB & K CB   & Q CB   & K VB  & $\mathrm{\Gamma}$ VB \\\hline
X$_{11}$      &    8.11     &  13.54    &  0.00   &  1.64  & 7.26    &    11.59  &  0.00   &  1.02 &      5.56   &    14.41  &  0.00   &  3.06&      5.50   &    14.52  &  0.00   &  2.04 \\
M$_{1}$      &     83.98   &   33.13   &  0.20   &  23.93 & 85.28  &   33.95   &  0.40   &  25.3 &      88.89  &    33.40   &  0.20   &  25.31 &      89.00  &    33.61   &  0.40   &  21.20   \\
X$_{12}$      &   7.91      &  16.25    &  0.00   &  20.86 & 7.46    &   15.32   &  0.00   & 20.41 &     5.56    &    15.24    &  0.00   &  17.96 &     5.50    &    15.98    &  0.00   &  17.18 \\
X$_{21}$      &   0.00      &    9.79   &  10.69  &  22.70 & 0.00 &    9.94   &   9.88 &    21.84&   0.00      &   9.39      &   12.52  &    19.59 &   0.00      &   9.13      &   11.94  &    18.40     \\
M$_{2}$      &  0.00      &   19.79   &  78.23  &  28.22 & 0.00 &    22.15  &  79.44  &  30.0&  0.00      &    19.42  &    75.55   &   30.00&  0.00      &    18.88  &    75.30   &   32.31  \\
X$_{22}$      &  0.00       &  7.50      & 10.89   &   2.66 & 0.00 &     7.03    & 10.28   &  1.43&  0.00       &     8.14    & 12.73 &    4.08 &  0.00       &     7.88    & 12.35 &    2.8 \\
\hline $d_z$ (e nm)      &  0.307       &  0.067      & -0.306   &   -0.018&  0.323       &  0.069      & -0.321   &   -0.019&  0.309       &  0.081      & -0.311   &   -0.022&  0.324 &  0.091      & -0.322  &   -0.042\\\hline\hline

\end{tabular}%
}
\end{table}

\begin{table}[ht]
\caption{Vacuum energy difference across MX$'$ TMD bilayers, $\Delta^P$(MX$'$), together with the parametrisation of the general configuration dependence of the potential drop, $\Delta^P(\mathbf{r}_0,d)$, Eq. \eqref{Eq:potential_jump}, calculated using the QE code. The value in parantheses for WSe$_2$ is that found above using the VASP code, with only $\sim$10\% variation between different codes in the prediction for $\Delta^P$.}
\begin{tabular}{ccccc}
  &$\Delta^P$(MX$'$) (meV) & $\Delta_a$ (meV)& $q$ (\AA$^{-1}$) & $d_0$ (\AA)\\ \hline \hline
{MoS$_2$}  &   69      & 12    & 2.22  &   6.5  \\
{MoSe$_2$} &   67      & 13    & 2.05  &   6.8    \\
{WS$_2$}   &   63      & 11    & 2.26  &   6.5\\
{WSe$_2$}  &   66(61)     & 10    & 2.10  &   6.9 \\\hline \hline
\end{tabular}
\label{tab_potential drops_fit}
\end{table}

\end{document}